\documentclass[twoside,twocolumn,9pt]{article}

\usepackage{extsizes}
\usepackage[left=1.5cm, right=1.5cm, top=1.785cm, bottom=2.0cm]{geometry}

\usepackage[utf8]{inputenc}
\usepackage{amsmath,amssymb}
\usepackage{graphicx}
\usepackage{hyperref,doi}
\usepackage[numbers,sort&compress]{natbib}

\def\vflow{v_{\textrm{flow}}}
\def\wall{\mathrm{wall}}
\DeclareMathOperator{\erfc}{erfc}
\def\vecr{\mathbf{r}}
\def\Pe{\mathrm{Pe}}
\def\Re{\mathrm{Re}}
\def\Ma{\mathrm{Ma}}

\def\acx{x}
\def\acy{y}
\def\acz{z}
\def\acr{r}
\def\yshift{y_\mathrm{shift}}
\def\Lbuffer{L_\mathrm{buffer}}
\def\vsound{v_\mathrm{sound}}

\def\torque{\mathcal{T}}
\def\com{\mathrm{com}}

\begin{document}

\author{Laurens Deprez\footnote{ORCID:\protect\href{http://orcid.org/0000-0003-4488-1615}{0000-0003-4488-1615}} \and Pierre de Buyl%
\footnote{Corresponding author} %
\footnote{ORCID:\protect\href{http://orcid.org/0000-0002-6640-6463}{0000-0002-6640-6463}}%
}
\date{Instituut voor Theoretische Fysica, KU Leuven, 3001 Leuven, Belgium}

\title{Passive and active colloidal chemotaxis in a microfluidic channel:\\mesoscopic and
  stochastic models}

\maketitle

\begin{abstract}
Chemotaxis is the response of a particle to a gradient in the chemical composition of the
environment.
While it was originally observed for biological organisms, it is of great interest in the
context of synthetic active particles such as nanomotors.
Experimental demonstration of chemotaxis for chemically-powered colloidal nanomotor was
reported in the literature in the context of chemo-attraction in a still fluid or in a microfluidic channel where
the gradient is sustained by a specific inlet geometry.
In this work, we use mesoscopic particle-based simulations of the colloid and solvent to
demonstrate chemotaxis in a microfluidic channel. On the basis of this particle-based model,
we evaluate the chemical concentration profiles in the presence of passive or chemically
active colloids, compute the chemotactic force acting upon them and propose a stochastic
model that rationalises our findings on colloidal chemotaxis.
Our model is also able to explain the results of an earlier simulation work that uses a
simpler geometry and to extend its interpretation.
\end{abstract}

\section{Introduction}

Nanomotors are nano- to micro-meter sized machines that use a local (internal or found in
the close environment) source of energy to move or perform work.
The catalytic conversion of a chemical fuel on catalytically coated colloids was used with
success for different types of motors (metallic rods~\cite{paxton_et_al_nanorods_2004},
polystyrene~\cite{howse_et_al_prl_2007} and silica~\cite{ke_et_al_jpc_a_2010} Janus
particles, dimers~\cite{valadares_el_al_sphere_dimers_small_2010}).
Chemically powered nanomotors represent very promising devices for the execution of tasks
such as sensing or cargo delivery in nano- to micro-meter scaled
environments~\cite{kapral_perspective_jcp_2013,wang_nanomachines_2013,ebbens_opinion_2016}.
Understanding the response of nanomotors to chemical concentration gradients, {\em
  chemotaxis}, is critical to engineer these possible applications.

The experimental characterisation of chemotactic motion can proceed either via chemical
sources that supply the environment with fuel, at specific ``target'' locations, or via a
flow that allows the sustainment of the gradient.
The first idea was used by Hong {\em et al} to demonstrate the occurrence of chemotaxis for
rod nanomotors~\cite{hong_chemotaxis_2007}.
The second strategy was used by Baraban {\em et al} in Ref.~\cite{baraban_anie_2013} where
the authors studied experimentally the chemotactic motion of chemically powered nanomotors
in a microfluidic channel and observed a lateral deviation of the
nanomotors when the fuel is input asymmetrically at the inlet of the cell.
The successful modeling and understanding of colloidal chemotaxis is also relevant for
biological systems, with the traditional example being the chemotactic behaviour of
bacteria~\cite{berg_chemotaxis_1975}.
The focus on biological chemotaxis has been put forward by Sengupta {\em et al} who also
used a microfluidic channel to observe the enhanced migration of enzymes across a
microfluidic channel in the presence of a substrate gradient~\cite{sengupta_jacs_2013}.
More recently, the sorting of enzymes on the basis of their chemotactic response has been
demonstrated in Ref.~\cite{dey_chemotactic_2014}.

Numerical simulations of chemotaxis can either assume an expression for a ``chemotactic
force'' or reproduce the chemotactic process itself, using a direct simulation of the
solvent. This article starts with the latter approach, using Molecular Dynamics simulation
with an explicit solvent in which several chemical species are represented.
We then formulate a continuous picture for the solvent concentration field and use it to
build a stochastic model that captures the chemotactic behaviour and its connection to our
simulation parameters.

The literature on particle-based modeling of chemotaxis for nanomotors is rather
scarce. Chen {\em et al}~\cite{chen_chemotactic_dimer_2016} considered a simplified system
in which a constant chemical gradient is imposed by the boundaries of the simulation cell.
In this work, we seek to imitate the experimental setup of Baraban {\em et
  al}~\cite{baraban_anie_2013} and to improve the theoretical understanding of this
experimentally relevant configuration for chemically powered nanomotors, albeit using a
simpler dimer-type nanomotor instead of the Janus and tubular microjet motors in the
experiment.
Our computational experiments builds gradually by starting with a non-catalytic spherical
colloid, then adding a catalytic property to the spherical colloid and finally using the
dimer nanomotor.
We come back to the results of Chen {\em et al}~\cite{chen_chemotactic_dimer_2016} and
extend their conclusions.

In section~\ref{sec:sim}, we review the mesoscopic simulation model.
The simulational implementation of the experiment of Ref.~\cite{baraban_anie_2013} is laid
out in section~\ref{sec:cell}.
The continuous representation of the fluid's diffusion profile, including in the situation
where chemical reactions occur on the surface of the colloid, is given in
section~\ref{sec:density}. There, we also compute the chemical gradient induced forces on
the colloids.
The stochastic model for the colloids is presented in section~\ref{sec:langevin}.
We give the results for both modeling strategies in section~\ref{sec:results} and conclude
in section~\ref{sec:conclusions}.
In appendix~\ref{sec:repro}, we provide full information on how to access the numerical code to
reproduce our findings.

\section{Simulation model}
\label{sec:sim}

The solvent consists of point particles with a mass $m_i$, a position $r_i$ and velocity
$v_i$. There is no force acting between solvent particles, their interaction is instead
modeled via cell-wise collisions at fixed time intervals $\tau$ using the Multiparticle
Collision Dynamics (MPCD) collision rule introduced by Malevanets and
Kapral~\cite{malevanets_kapral_mpcd_1999,malevanets_kapral_mpcd_2000}.
MPCD has been used successfully to investigate the dynamics of colloids in microfluidic
channels by Prohm {\em et al}, in the context of inertial
focusing~\cite{prohm_inertial_2012}, or by Nikoubashman {\em et al} to study the flow of
colloids in the presence of obstacles~\cite{nikoubashman_microfluidic_colloid_2013}, for
instance.

The evolution of fluid particles, in the presence of forces due to colloids or to the
flow-inducing field, is resolved numerically using the velocity Verlet
algorithm~\cite{allen_tildesley_1987,whitmer_luitjen_2010}
\begin{equation}
r_i(t + dt) = r_i(t) + v_i dt + \left(g + f_i(t) \frac{dt^2}{2 m_i}\right) ~,
\end{equation}
\begin{equation}
v_i(t+dt) = v_i(t) + \frac{dt}{2 m_i} \left( f_i(t) + f_i(t+dt) + 2 m_i g \right) ~,
\end{equation}
where $f_i$ is the total of pair forces on particle $i$ and $g$ is the external
acceleration.
The timestep $dt$ for MD is a fraction of the one for MPCD
\begin{equation}
dt = \frac{\tau}{N_\textrm{MD}} ~,
\end{equation}
where $N_\textrm{MD}$ is the number of MD steps between successive MPCD collisions.

At fixed time intervals $\tau$, the fluid particles are sorted in a lattice of cubic cells
of side $a$, whose origin is shifted randomly to ensure Galilean invariance~\cite{ihle_kroll_srd_2001}, and their
velocities are collided cell-wise according to
\begin{equation}
v_i' = v_\xi + \Omega_\xi \left( v_i - v_\xi\right)
\end{equation}
where the prime denotes the post-collision value, $\xi$ is the cell containing the particle
$i$, $\Omega_\xi$ a rotation operator of angle $\Theta$ in $\mathbb{R}^3$ around a randomly
chosen axis and $v_\xi$ is the centre-of-mass velocity of the cell.
The transport properties of a MPCD fluid can be computed analytically~\cite[and references
therein]{gompper_et_al_adv_polym_sci_2008,kapral_adv_chem_phys_2008}.
The simulation parameters are given in table~\ref{tab:cell-params}.

\begin{table}[h]
\centering
\caption{Simulation parameters for the chemotactic cell. The parameter files for reproducing the simulations are available publicly, see appendix~\ref{sec:repro}.}
\label{tab:cell-params}
\begin{tabular}{l l l}
  \hline
  Parameter & symbol & value\\
  \hline
  \hline
  Number density & $n$ & 10\\
  Solvent particle mass & $m$ & 1\\
  MPCD fluid density & $\rho$ & 10\\
  Cell size & $a$ & 1\\
  MPCD time step & $\tau$ & 0.5\\
  MD steps per MPCD step  & $N_{MD}$ & 50\\
  Cell dimensions & $L_x, L_y, L_z$ & 90, 60, 15\\
  Buffer length & $\Lbuffer$ & 20\\
  Temperature & $k_BT$ & $1/3$\\
  Acceleration & $g$ & 0.001\\
  MPCD collision angle & $\Theta$ & 2.6\\
  Reaction probability & $p$ & 1\\
  Colloid radius & $\sigma$ & 3\\
  Unit interaction energy & $\epsilon_A$ & 1\\
  \hline
\end{tabular}
\end{table}

One technique to obtain a Poiseuille flow is to impose a constant acceleration field in the
simulation cell~\cite{koplik_poiseuille_1988}, in combination with periodic boundary
conditions in the direction of the flow ($x$) and stick boundary conditions in the direction
transverse to the flow ($z$).
In Refs.~\cite{allahyarov_gompper_mpcd_flows_2002,whitmer_luitjen_2010}, this technique was
applied to MPCD fluids.
In the $y$ direction we use specular boundary conditions for the fluid, so that the flow
velocity profile $v_x(z)$ bears no dependence on $y$.
We implement stick boundary conditions for the flow in the $z$ direction with a combination
of bounce-back collisions and ghost particles at the boundary MPCD
cells~\cite{lamura_mpcd_epl_2001}. The ghost particles also set the temperature at the wall.
In this work the walls also provide a sufficient thermostatting action to compensate for the
flow-induced heating and no bulk thermostatting is used.
We have verified that the temperature of the fluid remains stable after a transient period
at a value that is about 3\% in excess of the temperature set at the walls.

The colloids evolve according to the velocity Verlet algorithm, similarly to the solvent
particles, but do not participate in the collisions and are not subject to the acceleration
field $g$.
In the case of dimer nanomotors, the distance between the two spheres is held constant using
the RATTLE algorithm~\cite{andersen_rattle_1983}.

The coupling with solvent particles is done via the shifted and truncated Lennard-Jones 12-6
potential, of the form
\begin{equation}
V_{ij} = \left\{
\begin{array}{l l}
4 \epsilon_{ij} \left( \left(\frac{\sigma_{ij}}{r_{ij}}\right)^{12} - \left(\frac{\sigma_{ij}}{r_{ij}}\right)^{6} + \frac{1}{4} \right) & \textrm{for } r_{ij} \le \sigma_{ij}\times 2^{1/6} \\
0 & \textrm{else}
\end{array}\right.
\end{equation}
where $\epsilon_{ij}$ denotes the strength of the potential and $\sigma_{ij}$ the radius of
the colloid. $i$ and $j$ represent the species of the solvent and colloidal particles and
$r_{ij}$ is the distance between them.
The variation of $\epsilon_{ij}$, depending on what type of solvent and colloid interact,
leads to a net force in the presence of chemical concentration gradients. This will be made
explicit in section~\ref{sec:density}.
$\epsilon_{ij}$ for interactions with the solvent of type $A$ is set to $1$ and defines the
energy scale. In the following, all quantities are expressed in simulation units with length
$a$ of the unit cell, mass $m$ of the solvent particles, energy $\epsilon_A$ and time
$\sqrt{a^2 m / \epsilon_A}$.
As the interaction between all colloids and solvent particles of species $A$ is equal, the
first subscript to $\epsilon_{\kappa,A}$ is dropped in the following.

Confinement of the colloids in the $z$ direction is obtained by a purely repulsive
Lennard-Jones 9-3 potential, of the form
\begin{equation}
V_{i} = \epsilon_{\wall, i} \left\{\frac{3\sqrt{3}}{2} \left( \left(\frac{\sigma_{\wall, i}}{z_{i}}\right)^{9} - \left(\frac{\sigma_{\wall, i}}{z_{i}}\right)^{3}\right) + 1 \right\} ~,
\end{equation}
where $z$ is here the distance to the closest horizontal wall.
The same potential is applied in the $y$ direction to avoid hypothetical crossings of the
lateral wall during a simulation.

MPCD fluids are well suited to describe chemical reactions influenced by catalytic
surfaces~\cite{ruckner_kapral_prl_2007} or in the bulk~\cite{rohlf_et_al_rmpcd_2008}.
At the surface of a catalytic colloid $C$, the reaction
\begin{equation}
A + C \to B + C
\end{equation}
converts fluid particles of type $A$ (the fuel) to fluid particles of type $B$ (the
product).
The reaction occurs when the fluid particles crosses the interaction region of the colloid
and is executed with a probability $p\in[0,1]$, when the fluid particles exits the interaction
region to avoid any discontinuity in the energy~\cite{ruckner_kapral_prl_2007}.

All simulations were performed with the open-source RMPCDMD
software~\cite{de_buyl_rmpcdmd_2017,rmpcdmd_1.0}.

\section{Cell design}
\label{sec:cell}

The inspiration for the design of the microfluidic cell comes from the experimental work on
the chemotaxis of nanomotors by Baraban {\em et al}~\cite{baraban_anie_2013}.
There, a channel is fed with fluid at a fixed flow rate via three inlets. As only an inlet
contains a chemical species of interest, i.e. the fuel for the nanomotors, the lateral
distribution of species is inhomogeneous.
The effect of the resulting gradient is a chemotactic behaviour that is observed by
monitoring the deviation angle of the nanomotors with respect to a straight line motion.

\begin{figure}[h]
\centering
\includegraphics[width=\linewidth]{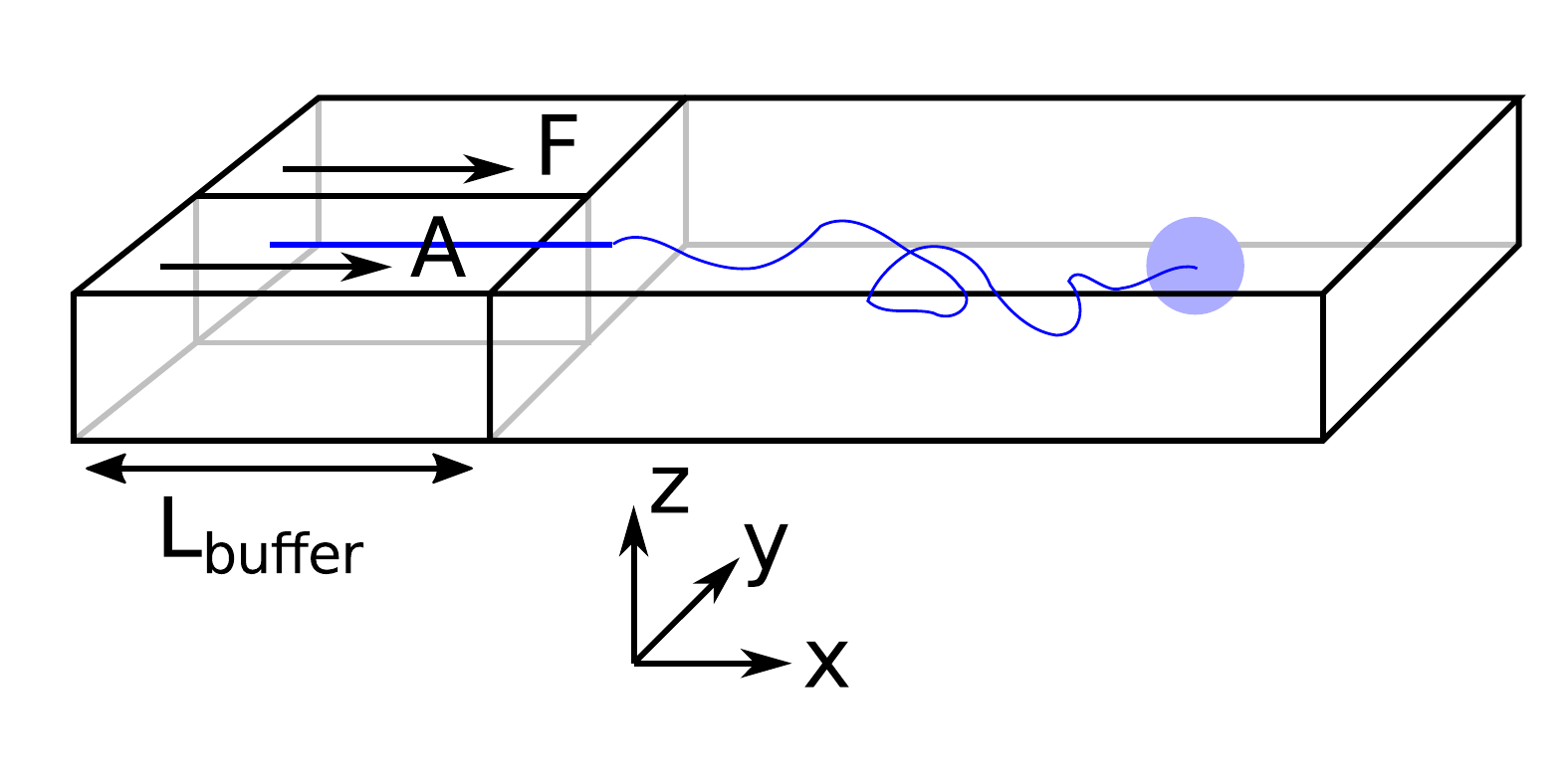}
\caption{Schematic description of the simulation cell for chemotaxis in a flow. There are
  two inlets on the left. In the buffer (leftmost region), solvent particles are set to
  species $A$ for $0<y<L_y/2$ and $F$ else. Lennard-Jones 9-3 potentials confine the
  colloids close to $z=L_z/2$. The colloid are initially placed in the $F$ inlet and
  constrained to a fixed $y$ and $z$ track. This constraint is released when $x>\sigma$.}
\label{fig:schema}
\vspace{1em}
\includegraphics[width=\linewidth]{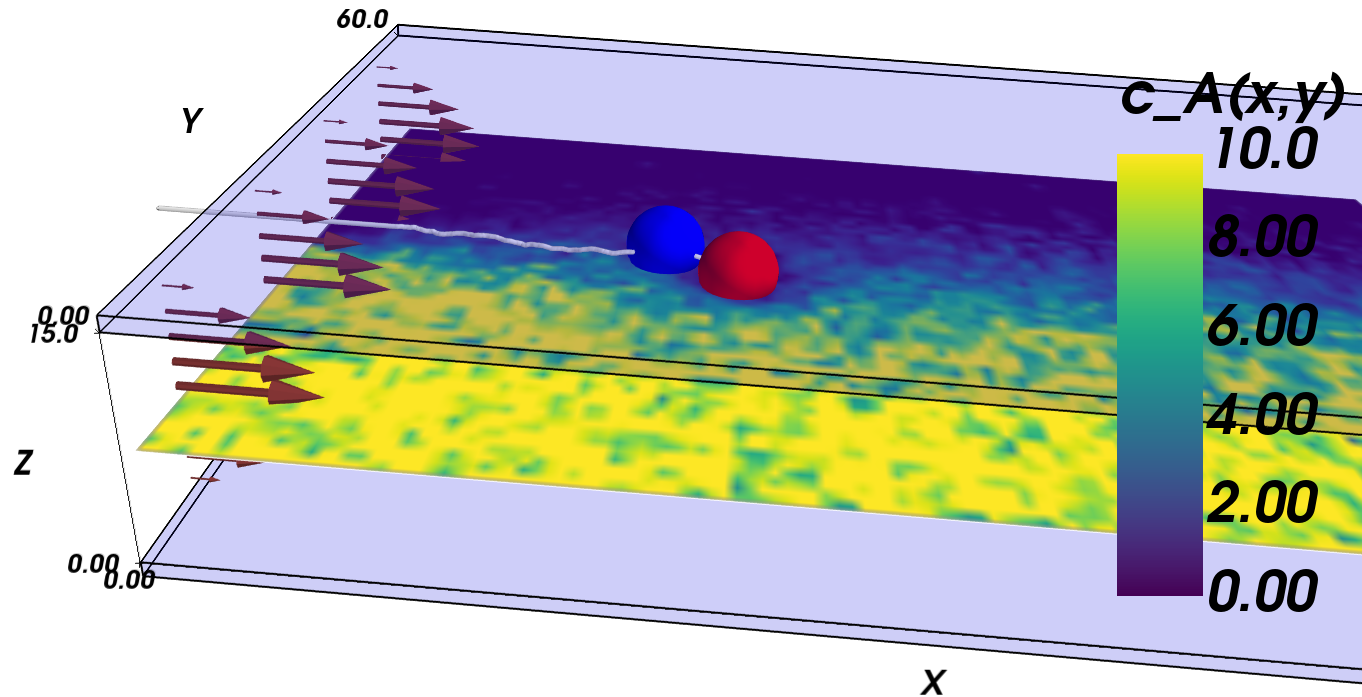}
\caption{Simulation snapshot for a dimer nanomotor simulation. The confining plates are made
  of transparent blue. The dimer is made of a catalytic (red) bead and of a non-catalytic
  (blue) bead and its centre-of-mass trajectory is shown as a white line. Here, the nanomotor
  is attracted toward the front of the cell (lower values of $y$) and displays a corresponding
  reorientation.
  The instantaneous concentration field of $A$ solvent particles around $z=L_z/2$ is shown
  in pseudocolor.}
\label{fig:snapshot}
\end{figure}

To reproduce the experimental features in a simulation, we have setup a thin channel with a
forced flow in the program \texttt{chemotactic\_cell} of the RMPCDMD software that is
illustrated in Fig.~\ref{fig:schema}.
A simulation snapshot in Fig.~\ref{fig:snapshot} shows the Poiseuille flow, the
concentration field for the fluid species $A$ and the trajectory of a dimer nanomotor.
To avoid reproducing the inlet channels at a large
computational cost, we assign the solvent species to be either $A$ (the fuel) or $F$ (a
neutral fluid species) arbitrarily, depending on the lateral position of the particles.
The total number of particles in the simulation is constant and particles are not created
nor destroyed but ``recycled'' when they re-enter the simulation cell due to the periodic
nature of the $x$ boundary.
The region of the cell where the forced attribution of species is applied is called the
buffer and is located in the region $0<x<\Lbuffer$.
In the following of the paper, the trajectories from the mesoscopic simulations are shifted
in $x$ by $-\Lbuffer$ to match the coordinates system of the stochastic model.

Although the simulations are three dimensional, the motion of the colloids is limited around
the centre of the cell in the z-direction by the confining walls.

The flow between two plates is of the Poiseuille type with maximal velocity $\vflow$ and
average velocity $v_{av}=2/3~\vflow$. We report the characteristic numbers of the flow in
Table~\ref{tab:fluid}.
For the work of Baraban~{\em et al}~\cite{baraban_anie_2013}, we use the values found in the
article for the flow rate (140$\mu$L per hour), the width (200$\mu$m per inlet, there are
three inlets) and temperature (300K). The height of the channel was confirmed by email to be
30$\mu$m. For the properties of water, we use reference data from the NIST~\cite[retrieved
January 11, 2017]{nist_thermophysical_webbook} (density
$\rho_\mathrm{water}=996.56$~kg/m$^3$, and viscosity
$\eta_\mathrm{water}=8.54~10^{-4}$Pa$\cdot$s) and take the diffusion coefficient
$D_{exp}=2~10^{-9}m^2/s$ that is a reasonable approximation for both water~\cite{wang_1965}
and hydrogen peroxide~\cite{kern_1954}.
For the MPCD fluid properties, we refer to the review by
Kapral~\cite{kapral_adv_chem_phys_2008}. The viscosity is
\begin{align}
\label{eq:eta}
  \eta = &\frac{k_BT\tau \rho}{2m} \left( \frac{ 5n - (n-1+e^{-n})(2-\cos\Theta-\cos 2\Theta) }{(n - 1 + e^{-n})(2-\cos\Theta-\cos 2\Theta)} \right)\cr
   &+ \frac{m}{18 a \tau} (n -1 + e^{-n})(1-\cos\Theta) ~,
\end{align}
and the self-diffusion coefficient is
\begin{equation}
D = \frac{k_B T \tau}{2m}\left(\frac{3n}{(n - 1 + e^{-n})(1-\cos\Theta)} - 1 \right) ~.
\end{equation}
We have verified that the velocity profile obeys the theoretical
prediction~\cite{allahyarov_gompper_mpcd_flows_2002}
\begin{equation}
\label{eq:v}
v_x(z) = v_{max} \frac{z (L_z-z)}{(L_z/2)^2} = \frac{\rho ~ g~L_z^2}{8\eta} \frac{z (L_z-z)}{(L_z/2)^2} ~.
\end{equation}

We have aimed for a similar fluid regime with respect to the experiment, that is a high
Péclet number ($\Pe$) flow, as our estimate for the concentration profile relies on
$\Pe\gg 1$ (see section~\ref{sec:density} and Ref.~\cite{ismagilov_diffusive_broadening}). A
value as high as for the experiment could not be obtained, but the important feature is that
the transport by the flow dominates the one by diffusion in the $x$ direction.
The Reynolds number $\Re$ should remain moderate for the laminar regime to hold.

The Péclet, Reynolds, and Mach number for the flow are computed as
\begin{equation*}
\Pe = L_z v_{av} / D ~,
\end{equation*}
\begin{equation*}
\Re = \vflow L_z / \eta ~,
\end{equation*}
and
\begin{equation*}
\Ma = \frac{v}{\vsound} ~,
\end{equation*}
where $\vsound = \sqrt{\frac{5}{3}\frac{k_BT}{m}}$ is the speed of
sound~\cite{prohm_inertial_2012}. The maximum Mach number is obtained for the maximum
velocity of the flow $v_{max}=\frac{\rho ~ g~L_z^2}{8\eta}\approx 0.095$ and is
$\Ma\approx 0.13$.

\begin{table}[h]
\centering
\caption{Characteristics of the microfluidic channel and of the fluid flow,
  both in the experiment of Ref.~\cite{baraban_anie_2013} and in our
  simulations.}
\label{tab:fluid}
\begin{tabular}{l | l l}
  \hline
  Number & Simulation & Experiment\\
  \hline
  \hline
  Width ($L_y$) & 60 & 600$\mu$m\\
  Height ($L_z$) & 15 & 30$\mu$m\\
  Average flow velocity $v_{av}$ & 0.063 & 2.16mm/s\\
  Maximum flow velocity $v_{max}$ & 0.095 & 3.24 mm/s\\
  $\Pe$ & 14 & 32\\
  Re & 0.48 & 0.076\\
  \hline
\end{tabular}
\end{table}

Besides the geometry of the cell, the simulation protocol differs slightly from its
experimental counterpart.
The colloid moves on a track at fixed $y=\frac{L_y}{2} + \yshift$ and
$z=\frac{L_z}{2}$. This restriction is lifted when the $x$ position of the colloid (centre
of mass) has passed $\Lbuffer$ plus its own radius. This allows for a systematic comparison
of the chemotactic drift across repeated runs without suffering from possible disturbance
from the resetting of particles in the inlets.
The shift $\yshift$ in the $y$ direction ensures that solvent particles in the interaction
range of the colloid are only of species $F$ and not influenced to a change of species in
the buffer region.

\section{Density profiles and surface interaction}
\label{sec:density}

In this section, we compute the stationary concentration field $c_\alpha(\vecr)$ for the
different chemical species in the cell and the resulting chemotactic force on the colloids.

We approximate the evolution of $c_\alpha(\vecr)$ by a 1D diffusion equation where the
spatial direction of the flow is proportional to time, i.e. $x=\vflow t$, and the diffusion
process acts on the transverse direction $y$.
This approximation has been tested experimentally in
Ref.~\cite{ismagilov_diffusive_broadening} and is only valid close to $z=L_z/2$. Close to
the boundaries $z=0$ and $z=L_z$ of the cell, a different transport regime is taking place.
Here, we neglect the variations in the $z$ direction for the computation of the
concentration $c_\alpha(\vecr)$ in the absence of catalytic particle.

Fixing $z=L_z/2$ and identifying the time and the $x$ coordinate results in
\begin{equation}
\label{2D-diff}
\partial_x c_\alpha(x,y) = \frac{D}{\vflow} \partial_y^2 c_\alpha(x,y)
\end{equation}

The separate inlets for the $A$ and $F$ chemical species translate into the initial
condition $c_A(0,y) = c_0 \Theta(L_y/2 - y)$, $c_B(0,y)=0$ and
$c_F(0,y) = c_0 \Theta(y - L_y/2)$ for $x=0$ (equivalently $t=0$).
The solution to Eq.~\eqref{2D-diff} is
\begin{eqnarray}
\label{sol-a}
c_A(x, y) &=& c_0 \left( 1 - \frac{1}{2} \erfc(\frac{{L_y}/{2} - y}{\sqrt{4 D x / \vflow}}) \right) \\
\label{sol-b}
c_F(x, y) &=& c_0 - c_A(x, y) \\
\label{sol-c}
c_B(x, y) &=& 0
\end{eqnarray}
where $\erfc$ is the complementary error function.
In the absence of a catalytically coated colloid, the value of $c_B$ remains zero at all times.
The average number density is a constant,
\begin{equation}
\label{rho}
\sum_\alpha c_\alpha(x,y) = \rho ~,
\end{equation}
in the low Mach number conditions here.

\begin{figure}[h]
\centering
\includegraphics[width=6cm]{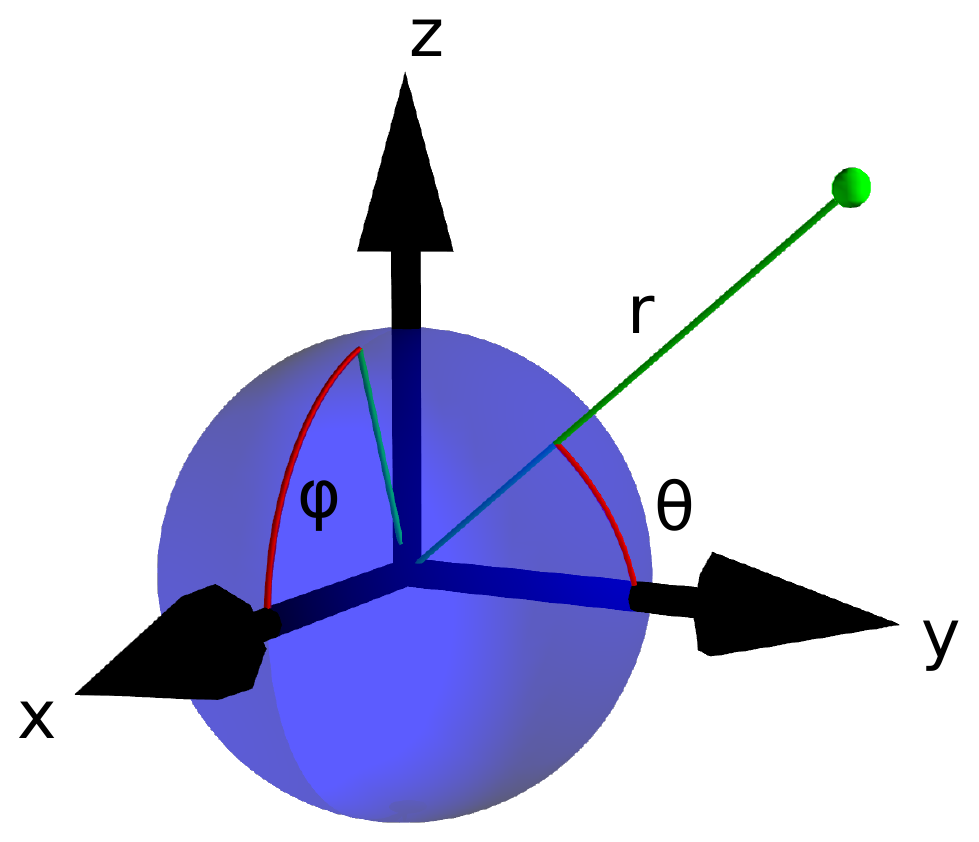}
\caption{The spherical coordinates centered on the colloid.}
\label{fig:coords}
\end{figure}

In the remainder of this section, we use coordinates centered around the colloid.
The spherical coordinates are defined by the following relations
\begin{equation}
\left\{
\begin{array}{l l}
\acx &= \acr \cos\varphi \sin\theta\cr
\acy &= \acr \cos\theta\cr
\acz &= \acr \sin\varphi \sin\theta
\end{array}\right.
\end{equation}
and are represented in Fig.~\ref{fig:coords}.

The presence of a catalytic colloid is taken into account in Eq.~\eqref{2D-diff} by a
radiation boundary condition (RBC) on the surface of the colloid, at radius $R$.
The boundary condition is applied at the limit of the interaction region,
i.e. $R=2^{1/6}\sigma$, where the continuum diffusive picture breaks down and where the
chemical reactions are triggered.
The RBC identifies the flux of chemicals with the consumption of the catalytic reaction, at
the surface of the colloid.
It was developed by Collins and Kimball~\cite{collins_kimball_1949} and rederived later by
other authors~\cite{schell_kapral_1981,gillespie_propensity_2009}.
The RBC at radius $R$ in the absence of external gradient, for the species $A$, is
\begin{equation}
\label{eq:rbc}
R k_D \partial_r c_A = k_0 c_A ~,
\end{equation}
where $k_D = 4 \pi R D$ is the diffusion-limited rate constant and
$k_0 = p R^2 \sqrt{8\pi k_BT /m}$ is the chemical rate~\cite{tucci_kapral_jcp_2004}.
$p$ is the reaction probability defined in section~\ref{sec:sim} and $m$ is the mass of a
fluid particle.

\subsection{Surface interaction}

Following the work of Rückner and Kapral~\cite{ruckner_kapral_prl_2007}, we sum the
interaction potential $V_{\kappa,\alpha}$ between a colloid of species $\kappa$ and the
fluid particles of species $\alpha$ over the interfacial region to obtain the potential
energy.
Differentiation with respect to the colloid's coordinate gives the force on the colloid:
\begin{equation}
\label{eq:force}
\vec F = \sum_\alpha \int d\vecr c_\alpha(\vecr) \vec\nabla V_{\kappa,\alpha}(\vec r) ~.
\end{equation}
In all the subsequent derivations, we rely on the solution of the diffusion equation in the
bulk of the fluid.
Within the interfacial region, the interaction with the colloid modifies the concentration
$c_\alpha$ by the Boltzmann weight~\cite{ruckner_kapral_prl_2007}. Equation~\eqref{eq:force}
then reduces to
\begin{align}
\vec F &= \sum_\alpha \int_{r = R} d\vecr \int dr c_\alpha(R \hat r) e^{-\beta V_{\kappa,\alpha}(r)} \vec 1_r \partial_r V_{\kappa,\alpha}(r)\cr
&= \sum_\alpha \int_{r = R} d\vecr c_\alpha(R \hat r) \vec 1_r \int dr e^{-\beta V_{\kappa,\alpha}(r)}  \partial_r V_{\kappa,\alpha}(r) ~,
\end{align}
where $\beta=\left(k_BT\right)^{-1}$ and $\vec 1_r$ is the unit vector pointing in the
radial direction.
Integrating by parts yields
\begin{equation}
\label{Lambdaforce}
\vec F = \frac{2}{\beta} \sum_\alpha \Lambda_{\kappa,\alpha} \int_{r = R} d\vecr c_\alpha(R \hat r) \vec 1_r~,
\end{equation}
where we have defined
\begin{equation}
\Lambda_{\kappa,\alpha} = \int_0^R dr r \left( e^{-\beta V_{\kappa,\alpha}(r)} - 1 \right) ~.
\end{equation}
Within this framework, it is sufficient to compute surface integrals over the colloids to
compute the force acting on them.
The use of Eq.~\eqref{Lambdaforce} in the literature is however limited to situations where
there is no external chemical gradient.

\subsection{Single passive colloid}

For a single passive colloid of type $N$, we use the solution \eqref{sol-a}-\eqref{sol-c}
together with Eq.~\eqref{Lambdaforce}.
\begin{equation}
\vec F_N =  \frac{2}{\beta} \left(\Lambda_{N,A}-\Lambda_{N,F}\right) \int_{r = R} d\vecr c_A(R \hat r) \vec 1_r~,
\end{equation}
Locally around the colloid, we use a first-order expansion of $c_\alpha$ in the lateral
direction.
\begin{align}
F_{N,y} &=  \frac{2}{\beta} \left(\Lambda_{N,A}-\Lambda_{N,F}\right) \int_{r = R} d\vecr \left( c_A(0, \theta) + \lambda R \cos\theta \right) \cos\theta\\
&=  \frac{8\pi}{3\beta} \left(\Lambda_{N,A}-\Lambda_{N,F}\right) \lambda R ~,\label{passive-Fy}
\end{align}
where $\lambda=\partial_y c_A(x,y)$
The explicit dependence of $F_{N,y}$ and $\lambda$ on the coordinates is omitted for brevity.

\subsection{Single active colloid}

To take into account the catalytic activity of a colloid, the diffusion
equation~\eqref{2D-diff} is solved with the radiation boundary condition (RBC).
Equation~\eqref{eq:rbc} expresses the flux that originates from the chemical activity of the
colloid, that is the only contribution to the flux in the situation where the RBC was
derived.
The total radial flux is the sum of the reaction-induced flux from Eq.~\eqref{eq:rbc} and of
the diffusive flux\footnote{The flux is multiplied by the surface of the sphere $4\pi R^2$
  as this is how the RBC is expressed in the literature.}
$4 \pi R^2 D \vec 1_r \cdot \vec \nabla c_A=Rk_D\lambda\cos\theta$.
\begin{equation}
\label{eq:rbc-flow}
R k_D \partial_r c_A = k_0 c_A + R k_D \lambda\cos\theta
\end{equation}
We make the following ansatz for the concentration field $c_A$:
\begin{equation}
\label{eq:cb}
c_A = c_0 + c_1 \frac{R}{r} + c_2 \left(\frac{R}{r}\right)^2 \cos\theta + \lambda r \cos\theta ~.
\end{equation}
Eq.~\eqref{eq:cb} is a solution of the diffusion equation and matches the boundary
condition~\eqref{eq:rbc-flow}.
The first three terms come from a truncation of the Legendre polynomial expansion of a
diffusion profile with a spherical catalytic sink.
The last term is needed to reflect the presence the external gradient.
We obtain the coefficients by inserting Eq.~\eqref{eq:cb} in Eq.~\eqref{eq:rbc-flow}:
\begin{equation}
\label{c0c1c2}
\left\{%
\begin{array}{l l l}
  c_0 &= c_A(x,y)\\
  c_1 &= - \frac{k_0}{k_0+k_D} c_0\\
  c_2 &= - \frac{k_0}{k_0+2k_D} \lambda R
\end{array}\right. ~,
\end{equation}
where we have used for $c_0$ the solution~\eqref{sol-a}. The value of $\lambda$ is obtained
by differentiating Eq.~\eqref{sol-a} with respect to $y$.

It is important to mention that (i) in the absence of a gradient, the coefficients in
Eq.~\eqref{c0c1c2} lead to the standard solution of a spherical sink and (ii) in the absence
of chemical activity, the solution~\eqref{sol-a} is recovered to linear order.

To obtain the solution for $c_B$, we observe that the {\em reactive flux} of $A$ particles
absorbed at the surface of the sphere equals the one of $B$ particles, except that the
currents flow in opposite directions. The solution to the diffusion equation with these
opposite flows on the surface is
\begin{equation}
\label{cb}
c_B = - c_1 \frac{R}{r} - c_2 \left(\frac{R}{r}\right)^2 \cos\theta ~.
\end{equation}

To compute the force in Eq.~\eqref{Lambdaforce}, we set $\Lambda_{C,F}=\Lambda_{C,A}$.
Using Eq.~\eqref{rho}, we find that the chemotactic force on the active colloid $C$ is
\begin{equation}
\label{eq:fcy}
F_{C,y} = - \frac{8\pi}{3\beta} \left(\Lambda_{C,B} - \Lambda_{C,A}\right) c_2
\end{equation}

\subsection{Dimer nanomotor}

A dimer nanomotor is made of two spheres linked rigidly, with a distance $d$ between their
centre of masses.
One ($C$) is catalytic and acts as a sink for $A$-species solvent particles and a source of
$B$-species solvent particles, creating locally a gradient centered around $C$.

\begin{figure}[h]
\centering
\includegraphics[width=.65\linewidth]{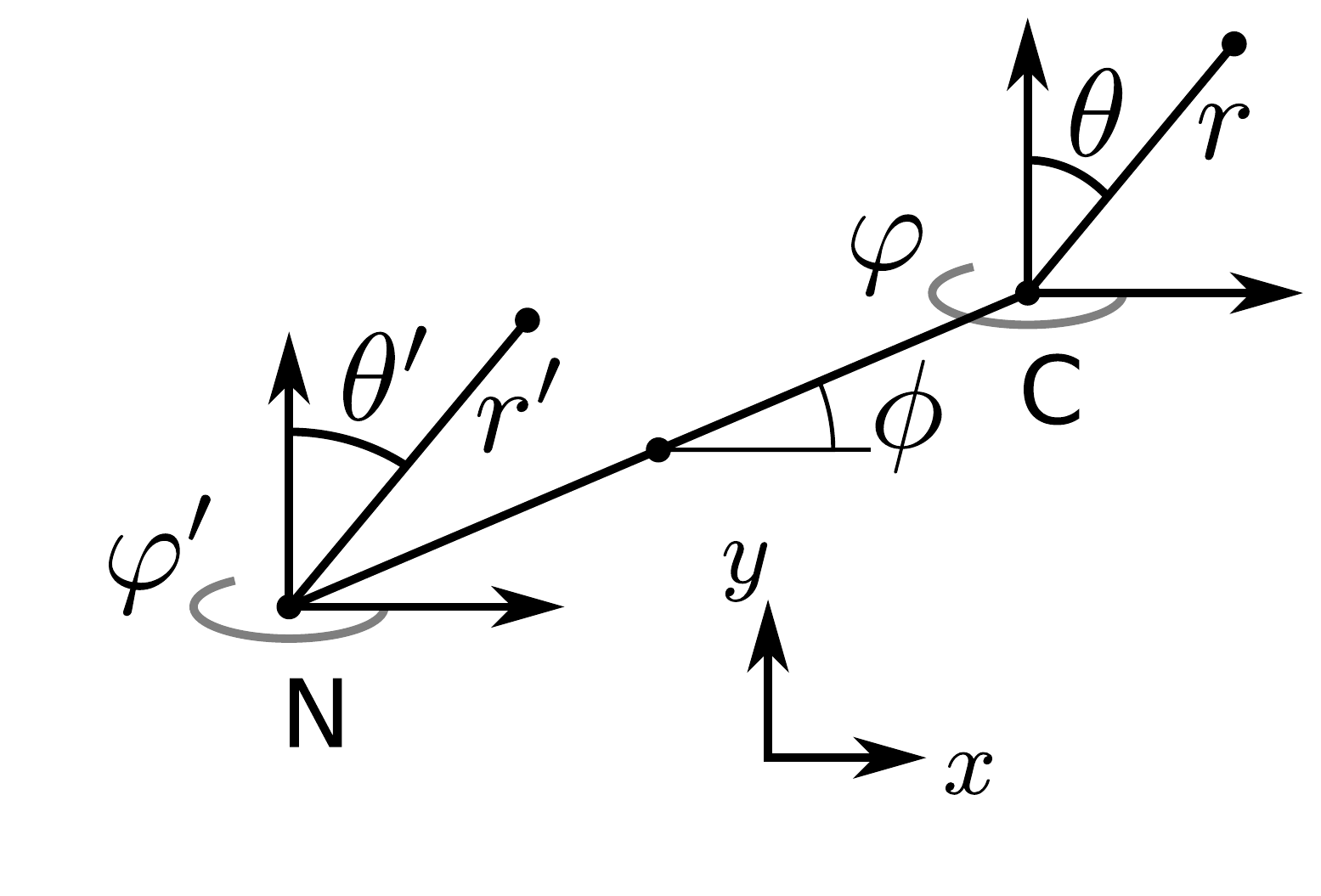}
\caption{Coordinate system for the dimer nanomotor.}
\label{fig:dimer}
\end{figure}

For the dimer nanomotor, the concentration fields depend on the presence of the catalytic
sphere of the motor in the same way as for the single active sphere and we reuse
Eqs.~\eqref{c0c1c2} and \eqref{eq:fcy}.
The force must be evaluated also on the non-catalytic sphere $N$ where the spherical
symmetry does not hold. We evaluate this expression numerically as
\begin{equation}
\vec F_{N} = \frac{2}{\beta}\sum_\alpha \Lambda_{N,\alpha} \int d\theta' \sin\theta' d\varphi' c_\alpha(R\hat r') \vec 1_{r'} ~,
\end{equation}
where the prime denotes the spherical coordinates around the $N$ sphere of the dimer,
illustrated in Fig.~\ref{fig:dimer}.

The forces on $C$ and $N$ are summed to obtain the centre-of-mass ($\com$) force:
\begin{equation}
\label{eq:f-com}
\left\{
\begin{array}{l l}
  F_{\com,x} &= F_{N,x}\\
  F_{\com,y} &= F_{C,y} + F_{N,y}
\end{array}\right. ~.
\end{equation}
The torque on the dimer is
\begin{align}
\label{eq:torque}
\torque &= \frac{d}{2} \left(\vec r_C - \vec r_\com\right) \wedge \vec F_C + \frac{d}{2} \left(\vec r_N - \vec r_\com\right) \wedge \vec F_N\cr
  &= d \hat u \wedge \left(\vec F_C - \vec F_N\right)
\end{align}

The motion of the dimer is studied via its centre-of-mass position in the $x$-$y$ plane and
its inclination $\phi$ with respect to the $x$ axis.

\section{Langevin dynamics for the colloids}
\label{sec:langevin}

For the spheres, the evolution of $x$ and $y$ (here, coordinates in the laboratory frame of
reference) is given by the overdamped Langevin equations~\cite{dhont_1996}.
\begin{align}
\label{langevin}
\dot x &= \vflow + \sqrt{2D} \xi_x\\
\dot y &= \frac{F_y(x/\vflow,y)}{\gamma} + \sqrt{2D} \xi_y
\end{align}
where $\vflow$ is the average velocity due to the flow at $z=L_z/2$, $\xi_x$ and $\xi_y$ are
normal distributed white noise, $D$ is the diffusion coefficient of the colloid and $\gamma$
its friction.
For the single spheres, we use $\gamma = 4\pi\eta\sigma$, the slip hydrodynamics friction,
and $D=k_BT/\gamma$.

For the dimer, we use the following Langevin equations for the centre of mass and orientation
\begin{align}
\label{langevin-dimer-1}
  \left(
  \begin{array}{l}
    \dot x - \vflow\\
    \dot y
  \end{array}\right)
  &=
  \left(
  \begin{array}{l l}
    \cos\phi & -\sin\phi\\
    \sin\phi & \cos\phi
  \end{array}\right)
  \left(
  \begin{array}{l}
    \frac{F_\parallel}{\gamma_\parallel} + \sqrt{2D_\parallel} \xi_\parallel\\
    \frac{F_\perp}{\gamma_\perp} + \sqrt{2D_\perp} \xi_\perp
  \end{array}\right)\\
  \label{langevin-dimer-2}
  \dot \phi &= \frac{\torque}{\gamma_r} + \sqrt{2D_r} \xi_\phi
\end{align}
where the projected forces $F_\parallel$ and $F_\perp$ are
\begin{equation}
\label{eq:projected}
  \left\{
  \begin{array}{l l}
    F_\parallel &= \left( F_{\com,x}\cos\phi + F_{\com,y}\sin\phi \right)\\
    F_\perp &= \left( -F_{\com,x}\sin\phi + F_{\com,y}\cos\phi \right)
  \end{array}\right. ~,
\end{equation}
and $\torque$ is the torque defined in Eq.~\eqref{eq:torque}.
The rotation operations in Eq.~\eqref{langevin-dimer-1} and \eqref{eq:projected} are due to
the difference in friction parallel and transverse to the axis of the dimer.

\begin{figure}[h]
\centering
\includegraphics[width=\linewidth]{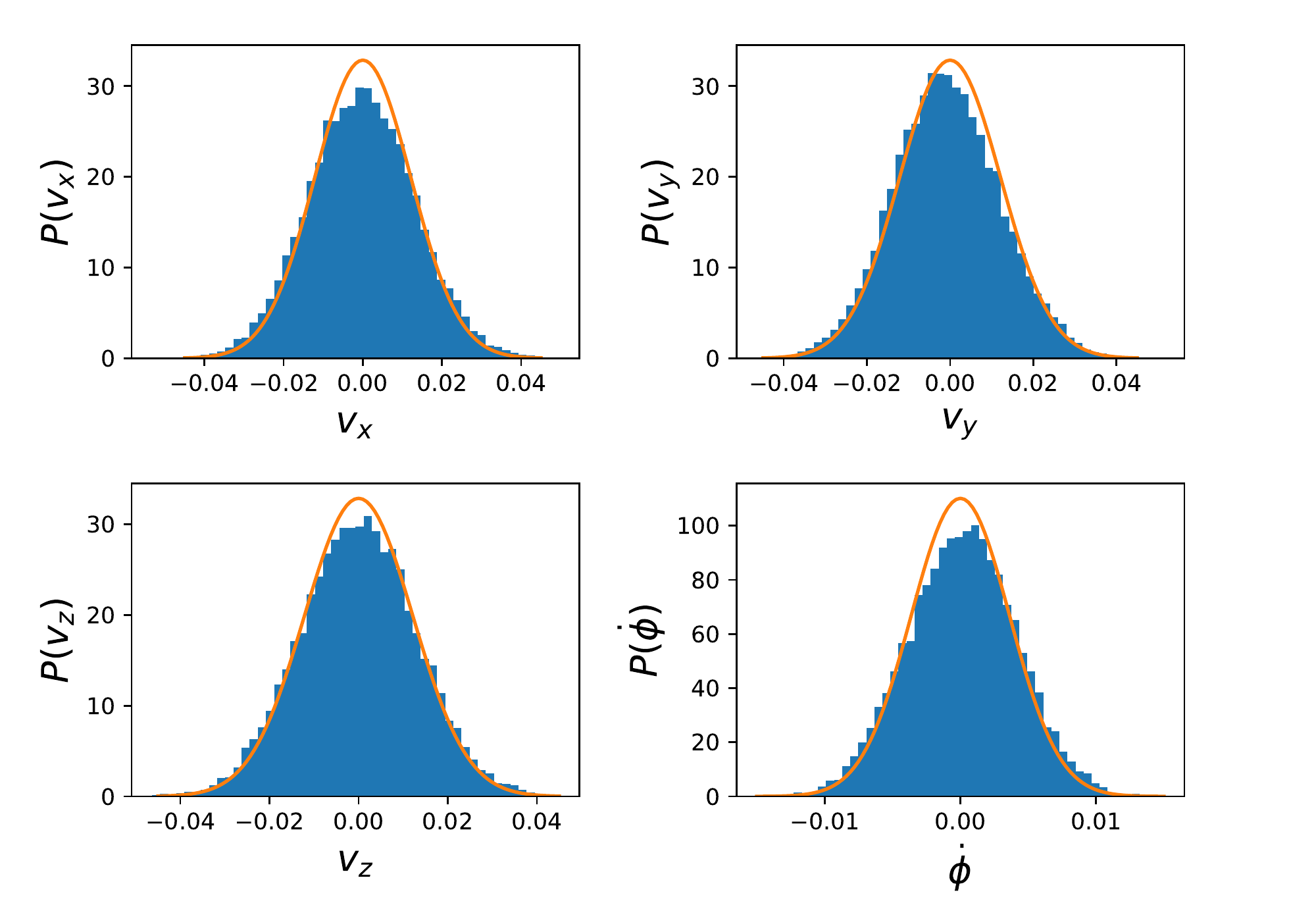}
\caption{The cartesian and angular velocity distributions for the dimer in the equilibrium
  simulations. The blue bars form an histogram over the simulation data and the full orange
  lines are the Boltzmann distributions at $k_BT=1/3$.}
\label{fig:dist}
\end{figure}

The diffusion coefficients for the dimers are obtained by performing equilibrium simulations
in a cell of dimension $\vec L = (32, 32, 15)$ and with no chemical reaction, all other
simulation parameters being taken equal.
The cartesian and angular velocity distributions for the dimer are shown in
Fig.~\ref{fig:dist}.
The dimer is an anisotropic colloid with axial symmetry. Accordingly, we compute separately
the diffusion coefficient parallel ($\parallel$) and transverse ($\perp$) to its axis as
\begin{equation}
D_\zeta = \int_0^\infty \langle \vec v_\zeta(\tau) \cdot \vec v_\zeta(0) \rangle d\tau ~,
\end{equation}
where $\zeta$ is $\parallel$ or $\perp$, following Ref.~\cite{dhont_1996}.
The projected velocities are defined by
\begin{equation}
\left\{
\begin{array}{l l}
  \vec v_\parallel &= \left(\vec v \cdot \hat u\right) \hat u\\
  \vec v_\perp &= \vec v - \vec v_\parallel
\end{array}\right. ,
\end{equation}
with $\hat u$ the unit vector in the direction joining the two spheres.
The angle $\phi$ measures the deviation of $\hat u$ in the x-y plane with respect to the
unit vector $\vec 1_x$.
The results of these simulations is $D_\parallel=2.0~10^{-3}$ and $D_\perp=1.5~10^{-3}$.
The rotational motion is characterised in the same fashion, using the velocity
autocorrelation of the angle $\phi$: $\langle \dot\phi(\tau) \dot\phi(0)\rangle$
with the result $D_r=1.4~10^{-4}$.
We determine the friction coefficient using the fluctuation-dissipation relation
\begin{equation}
D_\zeta = \frac{k_BT}{\gamma_\zeta} ~.
\end{equation}
The rotational time $D_r^{-1}\approx 7100$ is such that in the chemotaxis simulations the
dominant angular behaviour reflects the rotational drift. The contribution of rotational
diffusion is nevertheless visible in the spread of angles in Fig.~\ref{fig:nm_phi}.

\section{Results}
\label{sec:results}

\subsection{Passive sphere}
\label{sec:passive}

Here, we place a single sphere at the entry of the cell, in the upper y inlet surrounded by
solvent particles of species $F$.
When $\epsilon_A \ne \epsilon_{N,F}$, the y component of the force due to the chemical
gradient is non-zero.

\begin{figure}[h]
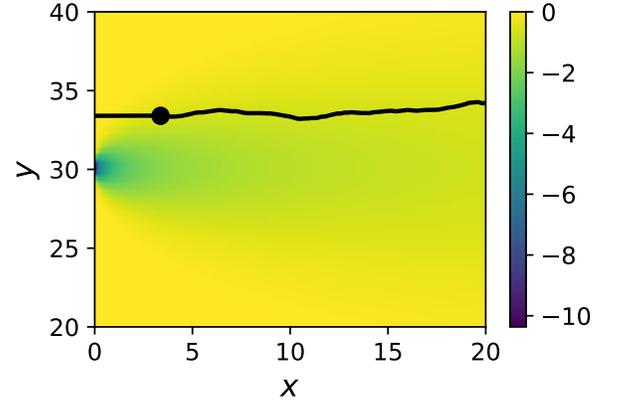

\centering
\includegraphics[width=.95\linewidth]{{{trajectory_and_gradient_1.00}}}
\caption{Example mesoscopic simulation of a passive sphere with $\epsilon_{N,F}=1$. The black
  line denotes the trajectory of the colloid, that starts on the left at the entrance of the
  cell and follows the flow to the right. The motion of the colloid is constrained to
  $y=L_y/2 + \yshift$ until the release point $x=\sigma$ (denoted by a black circle), after
  which it is influenced by the chemotactic force. The pseudocolor background indicates the
  strength of the gradient $\lambda(x,y) = \partial_y c_A(x,y)$.}
\label{fig:passive-sphere-example}
\end{figure}

A typical trajectory for the passive sphere is represented in the x-y plane in
Fig.~\ref{fig:passive-sphere-example}.
The main component is a displacement to the right under the influence of the flow, to which
thermal fluctuations are superimposed.
For this trajectory, $\epsilon_{N,F}=\epsilon_A$ so there is no chemotactic behaviour.

Upon changing $\epsilon_{N,F}$, a lateral force is exerted on the colloid due to the
combination of the asymmetry of $c_A$ and $c_F$ with respect to the colloid and of the
difference in surface interaction between the colloid and the $A$ and $F$ solvent species.
This situation is one of {\em passive} diffusiophoresis.

\begin{figure*}[h]
\centering
\includegraphics[width=.95\linewidth]{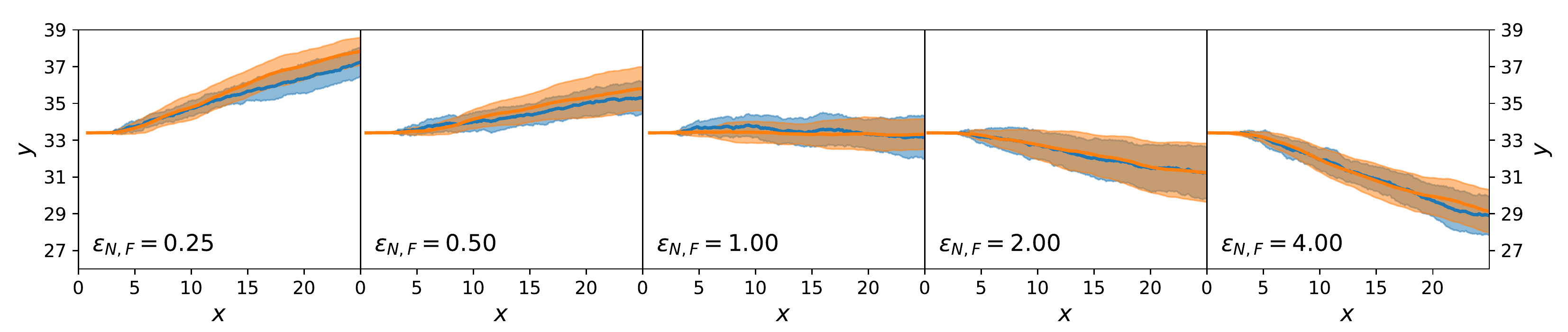}
\caption{Ensemble trajectories for the passive sphere simulations. The orange (blue) line is
  for the {\em average} position of the colloid in the mesoscopic (stochastic) simulations
  and the corresponding filled area indicates $\pm$ one standard deviation across
  realisations. There are 16 realisations for every set of parameters and simulation type.}
\label{fig:passive-sphere}
\includegraphics[width=.95\linewidth]{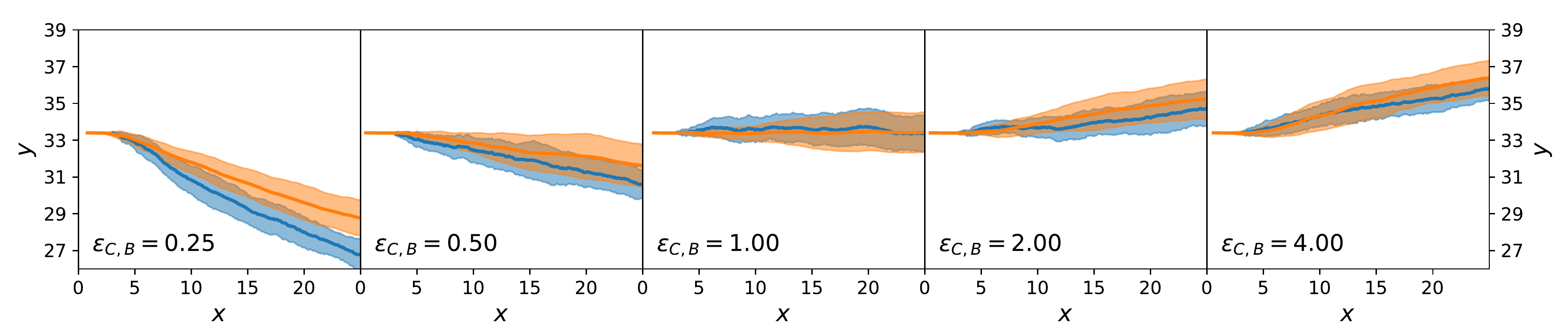}
\caption{Ensemble trajectories for the active sphere simulations.
  Organisation as in Fig.~\ref{fig:passive-sphere}.}
\label{fig:active-sphere}
\end{figure*}

The results for passive spheres are summarised in Fig.~\ref{fig:passive-sphere} for all
chosen values of $\epsilon_{N,F}$. For the central panel $\epsilon_{N,F}=\epsilon_A=1$, the
sphere moves in $x$ down the flow (i.e. from left to right in the figure) and undergoes
diffusive motion in $y$.
For $\epsilon_{N,F}<\epsilon_A$, we have $\Lambda_{N,A}-\Lambda_{N,F}<0$ and we expect from
Eq.~\eqref{passive-Fy} a lateral force whose sign is opposite to the gradient $\lambda$. As
$c_A(x,y)$ decreases for increasing $y$, $\lambda<0$ and the chemotactic force is upward, as
confirmed by the panels $\epsilon_{N,F}=0.25$ and $\epsilon_{N,F}=0.50$.
Conversely, for $\epsilon_{N,F}>\epsilon_A$, the chemotactic drift is negative. This is
shown for $\epsilon_{N,F}=2$ and $\epsilon_{N,F}=4$ in Fig.~\ref{fig:passive-sphere}.

Figure~\ref{fig:passive-sphere} reports the results for both the mesoscopic model and the
stochastic model. Given the fluctuating nature of the colloid dynamics, we have repeated the
simulations 16 times and denote by a shaded area one standard deviation below and above the
average results.
We observe that the average trajectories are similar and also that the spread of
trajectories is similar.

This first result on passive chemotaxis provides here the simplest context for the
experimental setup and allows a possible calibration of the quantity
$\Lambda_{N,A}-\Lambda_{N,F}$ without any dependence on reaction rates.

\subsection{Active sphere}
\label{sec:active}

We now turn to the chemotactic behaviour of an active sphere.
The injection setup is the same as for the the passive sphere but now
$\epsilon_{C,F}=\epsilon_A$ and the mechanism that creates a systematic lateral motion for
the passive sphere, that is proportional to $\Lambda_{C,A} - \Lambda_{C,F}$, is effectively
zero.

What occurs, instead, is that the local gradient in $c_A$ generates an asymmetric
distribution also for $c_B$ that, together with a nonzero value for
$\Lambda_{C,B} - \Lambda_{C,A}$, creates an original combination of passive and active
diffusiophoresis.
Using Eq.~\eqref{eq:fcy} and observing that $c_2>0$ (as $\lambda<0$ everywhere in the
simulation), we expect a downward chemotaxis for $\Lambda_{C,B} - \Lambda_{C,A}>0$
(i.e. $\epsilon_{C,B}<\epsilon_A$) and an upward chemotaxis for
$\Lambda_{C,B} - \Lambda_{C,A}<0$ (i.e. $\epsilon_{C,B}>\epsilon_A$).
We explore the effect of changing $\epsilon_{C,B}$ in Fig.~\ref{fig:active-sphere} that
confirms this direction for the chemotaxis of the active colloid.
As for the passive sphere, the comparison between the mesoscopic and the stochastic models
is positive.

Even though the situation of the active sphere is simpler than the experiments on nanomotors
of Ref.~\cite{baraban_anie_2013}, it already contains a complex ingredient: the chemotactic
force is entirely caused by the self-generated concentration field around the colloid.
Thanks to the derivations in section~\ref{sec:density}, we understand how the asymmetry of
the imposed concentration fields of $A$ and $F$ lead to an asymmetry in the concentration of
$B$ that gives rise to chemotaxis.

\subsection{Dimer nanomotor}
\label{sec:dimer}

For the dimer nanomotor, we track in the simulations the centre-of-mass position
$\vec r_\com$ and the orientation $\phi$.
The release from the injection track occurs here as soon as both spheres have exited the
buffer region and the initial orientation is $\phi=0$.

We have chosen, for the simulation parameters, that $\epsilon_{C,\alpha}=\epsilon_{N,\alpha}$,
for all solvent species $\alpha$.
This is at variance with simulations of the prototypical dimer
nanomotor~\cite{ruckner_kapral_prl_2007,tao_kapral_nanodimer_jcp_2008,valadares_el_al_sphere_dimers_small_2010,chen_chemotactic_dimer_2016}
where $\epsilon_{C,A}=\epsilon_{C,B}=\epsilon_A \neq \epsilon_{N,B}$.
Preliminary tests showed that in this situation the trajectories are pathological in the
sense that there was no ``gentle'' deviation of the orientation $\phi$ and that no
systematic tendency could be found.
For nanomotors in a bulk environment with no a priori gradient in the chemical
concentrations, this choice would be irrelevant.

\begin{figure*}[h]
\centering
\includegraphics[width=.95\linewidth]{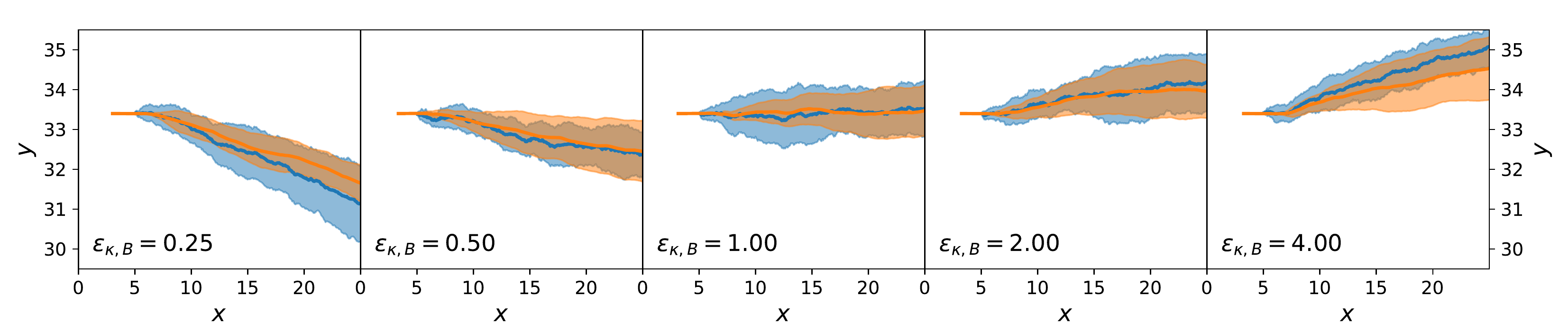}
\caption{Ensemble trajectories for the dimer nanomotor simulations. The centre-of-mass position is used.
  Organisation as in Fig.~\ref{fig:passive-sphere}.}
\label{fig:nm}
\includegraphics[width=.95\linewidth]{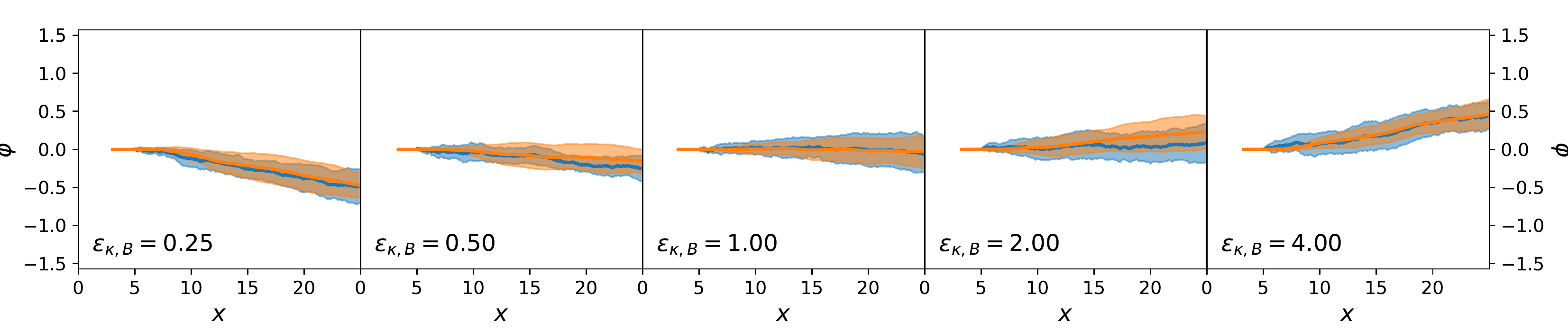}
\caption{Ensemble trajectories of the dimer nanomotor simulations. Here, the orientation $\phi$ with respect to the $x$ axis is shown.
  Organisation as in Fig.~\ref{fig:passive-sphere}.}
\label{fig:nm_phi}
\end{figure*}

Both the lateral deviation in the x-y trajectories and the angular deviation for $\phi$ are
reported in Figs.~\ref{fig:nm} and \ref{fig:nm_phi} and are consistent between the
mesoscopic simulations and the stochastic model.
This confirms the adequateness of the combined force-torque evolution given by
Eqs.~\eqref{langevin-dimer-1}-\eqref{langevin-dimer-2} as both the direction in $y$ and in
$\phi$ are reproduced.
The origin of the lateral deviation can be understood as coming not only from the
redirection of the self-propelling force, that is oriented along the dimer axis, but also
from the net lateral force on the centre of mass whose direction we can infer from
Eqs~\eqref{c0c1c2}, \eqref{eq:fcy} and~\eqref{eq:f-com}.
To our knowledge, this is the first time that the continuum force computation of
Ref.~\cite{ruckner_kapral_prl_2007} is extended to compute the lateral force on a nanomotor
and the resulting torque on the dimer nanomotor.

\subsection{Comparison to earlier simulation work}

Aside from the geometric considerations that are specific to the microchannel, the
stochastic model that we have designed can accommodate other situations.
We use it to test, in a qualitative sense, the observations of
Ref.~\cite{chen_chemotactic_dimer_2016}: we re-use the diffusion coefficients that we have
obtained with our cell geometry and fluid parameters but change the type of gradient and the
interaction type of the motor to match those of the earlier paper, which we call the
``constant gradient'' model.
In Ref.~\cite{chen_chemotactic_dimer_2016}, Chen {\em et al} design a cell in which the
gradient of the chemical species $A$ and $F$ is constant\footnote{For clarity, we reuse the
  species labels and orientation of the gradient used in the present work. This should be
  kept in mind when comparing with the work of Chen {\em et al} where the gradient is along
  x and the species are labelled $F$ for the fuel (here, $A$), $S$ for the inert fluid
  (here, $F$) and $P$ for the reaction product (here, $B$).}, i.e.
$c_A(x,y)=\rho \frac{y}{L_y}$ and $c_F(x,y)=\rho \frac{L_y-y}{L_y}$ in the absence of
chemical reaction, and the interaction parameters $\epsilon$ between the $C$ sphere and both
solvent species is identical, $\epsilon_{C,A}=\epsilon_{C,F}=\epsilon_{C,B}$, so that there
is no chemical gradient force on the $C$ sphere (explicitly,
$\Lambda_{C,A}=\Lambda_{C,B}=\Lambda_{C,F}$).

The main finding of Chen {\em et al} are
\begin{enumerate}
\item The average orientation of the nanomotor is toward $y=0$, even though by a small
amount (see inset of Fig.~4(c) of Ref.~\cite{chen_chemotactic_dimer_2016}).
\item The average trajectory of the nanomotor is toward $y=L_y$ (see Fig.~7(a) of
Ref.~\cite{chen_chemotactic_dimer_2016}).
\end{enumerate}
We will show here that these results depend strongly on the choice of parameters for the
interaction between the colloid surfaces and the solvent via three observations: the
distribution of trajectories $y(t)$, the distribution of angles $P(\theta)$, and the
distribution of position $P(y)$.

\begin{figure}[h]
\centering
\includegraphics[width=\linewidth]{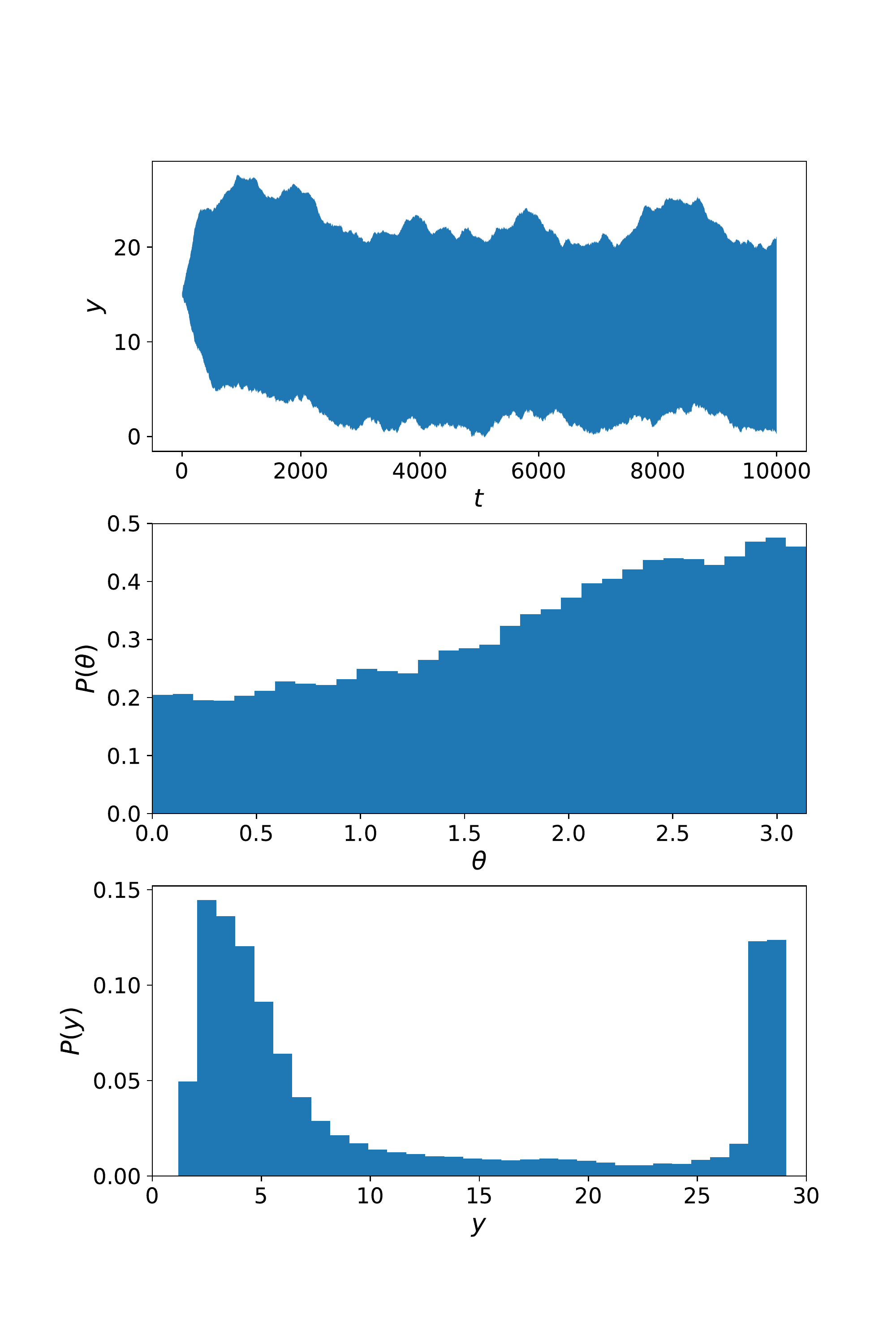}
\caption{Simulations of the constant-gradient setup with $\epsilon_{N,B}=0.25$. Averages are
  performed over 40 realisations for $2.5~10^4 < t <10~10^4$ (neglecting the start of the
  simulations for about 3 typical rotation times $1/D_r$). From top to bottom, the panels
  show the area between $\langle y(t) \rangle - \sigma_y(t)$ and
  $\langle y(t) \rangle + \sigma_y(t)$, the distribution of angles $\theta$ with respect to
  the axis $y$ and the distribution of $y$.}
\label{fig:cg_N}
\end{figure}

We have performed stochastic simulations with
$\epsilon_{\kappa,A}=\epsilon_{\kappa,F}=\epsilon_{C,\alpha}=1$ (for all solvent species
$\alpha$ and colloid species $\kappa$) and $\epsilon_{N,B}=0.25$, $L_y=30$, $L_z=15$. A
harmonic wall with spring constant $k_\wall=10$ prevents the exit of the colloid and is
turned on for $y<2$ or $y>\L_y-2$.
The nanomotors were started in the $x$ direction, as in
Ref.~\cite{chen_chemotactic_dimer_2016}, to avoid an a priori bias.
The results are displayed in Fig.~\ref{fig:cg_N}. The distribution of $y(t)$ trajectories in
the upper panel shows a large spread, from which we cannot conclude of a dominant
chemotactic behaviour. Examining $P(y)$, we see however that there is an average
accumulation of particles close to $y=0$.
The angular distribution $P(\theta)$ does show a bias toward $y=0$ (equivalently
$\theta=\pi$), as in Ref.~\cite{chen_chemotactic_dimer_2016}.

\begin{figure}[h]
\centering
\includegraphics[width=\linewidth]{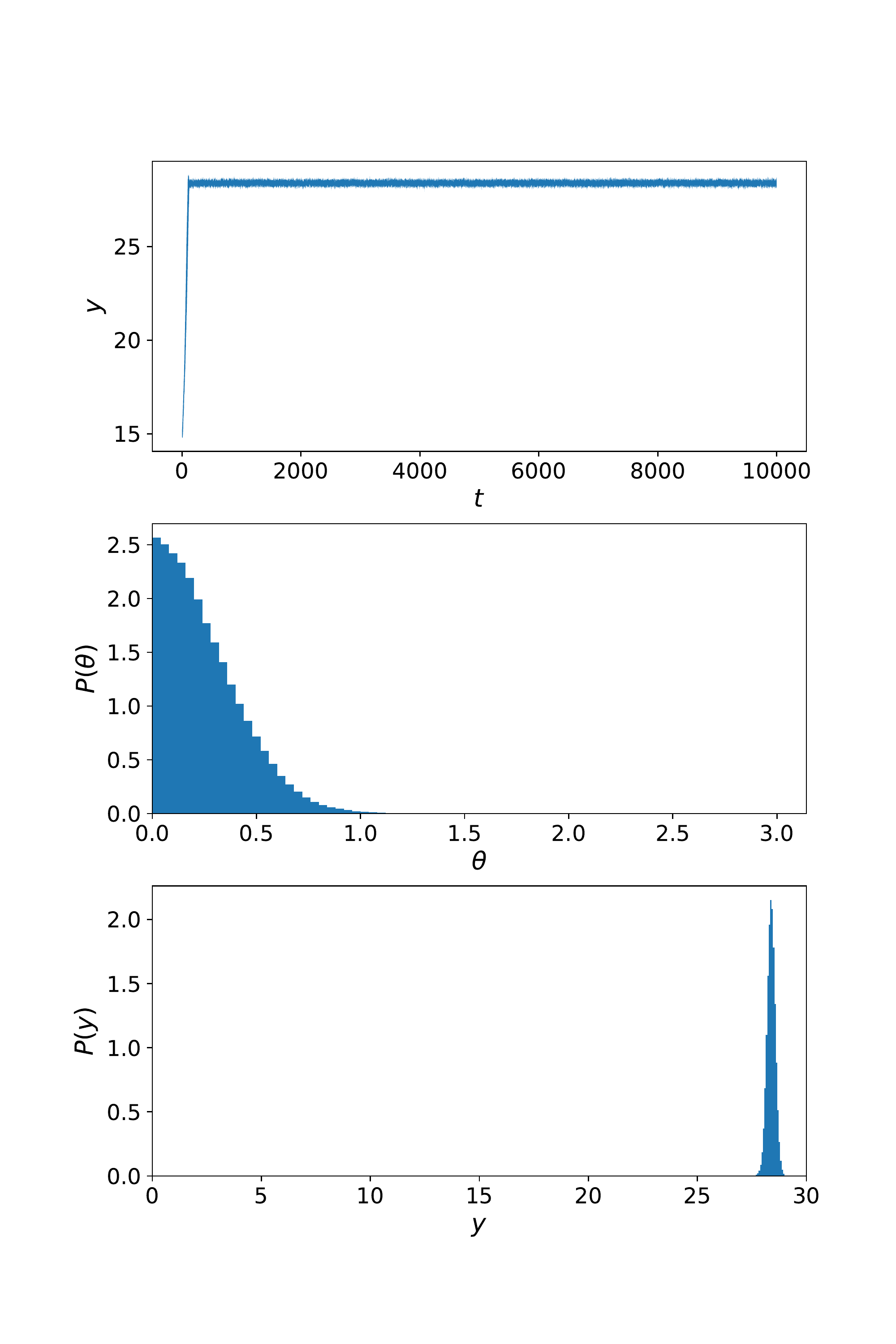}
\caption{Simulations of the constant-gradient setup with $\epsilon_{N,B}=0.25$ and
  $\epsilon_{C,B}=0.25$. Averages are performed over 20 realisations for
  $2.5~10^4 < t <10~10^4$. Panels as in Fig.~\ref{fig:cg_N}.}
\label{fig:cg_N_C}
\end{figure}

Until here, there was no chemically-induced force on the $C$ sphere. To assess the
importance of this choice, we perform another set of simulations
$\epsilon_{C,B}=\epsilon_{N,B}$ that are shown in Fig.~\ref{fig:cg_N_C}. This
parametrisation is the one used for our main results of sections~\ref{sec:passive},
\ref{sec:active}, and \ref{sec:dimer}.
From Eqs.~\eqref{eq:fcy}-\eqref{eq:torque}, we know that the torque on the nanomotor will be
influenced by this choice. This is reflected in the angular distribution $P(\theta)$ in
Fig.~\ref{fig:cg_N_C} that now displays a strong orientation toward $\theta=0$ (i.e. the
nanomotor is oriented toward $y=L_y$). As a result, there is no competition between a
downward-facing nanomotor and the greater velocity for upward-facing nanomotor orientation
that is observed in Ref.~\cite{chen_chemotactic_dimer_2016} and the distribution of $y(t)$
trajectories is narrow, showing a strong chemotactic behaviour. This is confirmed by the
distribution of positions $P(y)$.

To conclude this comparison, we have observed competing effects of orientation and
propulsion, leading to different chemotactic behaviours. The distribution of angles, for
instance, depends stronly on the parameters chosen for the colloid-fluid interaction:
For the interaction choice of Chen {\em et al},
$\epsilon_{C,A}=\epsilon_{C,F}=\epsilon_{C,B}$, we also find an average orientation toward
$y=0$, but for the other choice that we used the average orientation is toward $y=L_y$.
The overall chemotactic behaviour will however also depend on the diffusive and propulsive
properties of the dimer.

Although further research on the experimental characterisation of the surface properties of
the nanomotors is needed, our stochastic model provides an interesting tool to relate these
properties to the effective chemotactic behaviour of nanomotors.

\section{Conclusions}
\label{sec:conclusions}

We have proposed a particle-based simulation setup, based on Multiparticle Collision
Dynamics and Molecular Dynamics, to study the chemotactic motion of passive and active
colloids.
The concentration field that drives this motion is sustained by the flowing input of the
cell, via two inlets, as is done in the experiment of Ref.~\cite{baraban_anie_2013}.

We then constructed an approximate solution for the chemical concentration profile in the
microfluidic channel in the presence of an active colloid and built a stochastic model with
a microscopic expression for the systematic force.
This model provides a sound basis for the microscopic origin of chemotaxis, i.e. it explains
why the colloids move towards higher or lower values of the coordinate $y$.
Upon extending this model to a two-sphere dimer nanomotor, we also gain an understanding of
why the nanomotor changes its orientation via a systematic torque.
We have thus improved on the formal understanding of the chemotactic motion in the
microfluidic channel.

Our stochastic model, adapted to the setup with a constant concentration gradient by Chen
{\em et al} reproduces the observation that the nanomotor tends to orient opposite to the
gradient. While Chen {\em et al} observed a net positive chemotaxis, this is not the case
in the present work.
The reason is that the interplay between orientation and propulsion depends on the geometry
of the motor and on the choice of parameters.
This reasoning is supported by changing the surface interaction also for the $C$ bead,
resulting in a cooperation of orientation and propulsion leading to positive chemotaxis.
The specific change in $\epsilon_{C,B}$, that is not used in the simulation literature but
introduced here in subsection~\ref{sec:dimer}, allowed us to probe a regime of well-defined
positive chemotaxis and demonstrates that the class of simulation models introduced by
Rückner and Kapral~\cite{ruckner_kapral_prl_2007} possesses a very rich phenomenology.

It is interesting to note that our stochastic model requires only the input of the surface
interaction parameters $\Lambda_{\kappa,\alpha}$ and of the diffusion coefficients of the dimer.
In principle, it could be extended to other forms of the interaction potential or rely on a
characterisation of $\Lambda_{\kappa,\alpha}$ that does not reveal the full functional form
of this potential, and also to other motor geometries.
While we have chosen $\Lambda_{N,\alpha} = \Lambda_{C,\alpha}$ for all solvent species
$\alpha$ for this first investigation of the model, there is room for possible qualitative
changes with other choices, as we have witnessed for the constant gradient configuration.

The overall speed of the motor, as it was the case in earlier simulation studies for dimer
nanomotors, only depends on $\Lambda_{N,A}-\Lambda_{N,B}$~\cite{ruckner_kapral_prl_2007}. The
torque, on the other hand, depends also on $\Lambda_{C,A}-\Lambda_{C,B}$. A difference in
these quantities, and thus in the surface properties of both sides of the nanomotor, could
be revealed in a controlled manner in experiments and this reinforces the importance of the
microfluidic channel configuration for chemotactic studies.

The present work can be extended to other types of motors, notably the Janus nanomotors that
are used in Ref.~\cite{baraban_anie_2013} and for which colloidal assemblies have already
been used in mesoscopic simulations~\cite{de_buyl_kapral_nanoscale_2013}. This line of
research is promising to test {\em in silico} the behaviour of different motor geometries.
Using models suitable for the chemo-mechanics of enzymes, at a mesoscopic
level~\cite{cressman_et_al_pre_2008,echeverria_enzyme_pccp_2011,togashi_mikhailov_elastic_network_pnas_2007},
could provide very fruitful advances for understanding the recent works on enzyme
chemotaxis~\cite{sengupta_jacs_2013,dey_chemotactic_2014}, especially given the fact that
multiple inlets microfluidic devices originate from studies on bacterial
chemotaxis~\cite{mao_pnas_2003} and have been used for the enzyme
studies~\cite{dey_chemotactic_2014,sengupta_jacs_2013}.

\section*{Acknowledgements}

The authors thank Raymond Kapral for interesting discussions and Samuel Sanchez for
communicating the height of the cell in Ref.~\cite{baraban_anie_2013}.
PdB is a postdoctoral fellow of the Research Foundation-Flanders (FWO).

\appendix

\section{Computational reproducibility}
\label{sec:repro}

In this appendix, we review how the present work can be reproduced.
The software and parameter files are all available publicly under open-source licences.
We have prepared supplementary material that contains the relevant parameter files for the
mesoscopic simulations and the code for data analysis and archived them with
Zenodo\footnote{\url{https://zenodo.org/}}, available as
Ref.~\cite{colloidal_chemotaxis_companion_2017}.

All mesoscopic simulations are performed using the open-source software package
RMPCDMD~\cite{de_buyl_rmpcdmd_2017,rmpcdmd_1.0} for the simulations of passive and active
colloids, developed by the authors with Mu-Jie Huang and Peter Colberg.
The output of RMPCDMD consists of H5MD~\cite{h5md_cpc_2014} files that contain the full
trajectory for the colloids, the thermodynamic observables and the correlation functions
(velocity autocorrelation functions and mean-squared displacement).

All stochastic simulations are performed using Python, NumPy and Cython in a
Jupyter\footnote{\url{http://jupyter.org/}} notebook.
The analysis of both types of simulations and the execution of the stochastic model
simulations are done in the notebook \texttt{colloidal\_chemotaxis.ipynb}, except for the
constant gradient model that is implemented in a separate Cython module
\texttt{stochastic\_nanomotor.pyx} and driver program \texttt{run\_cg\_nm.py}.
The equilibrium simulations for the dimer are analysed in the notebook
\texttt{diffusion.ipynb}.

The references for the software are the following: NumPy~\cite{numpy_csie_2011} is used for
all numerical work in the analysis of the mesoscopic model and overall for the stochastic
model, SciPy~\cite{scipy-web} is used for computing the erf function and for numerical
integration, matplotlib~\cite{matplotlib_2007} to generate the figures,
Mayavi~\cite{mayavi_2011} for Fig.~\ref{fig:snapshot},
h5py~\cite{collette_python_hdf5_2013} to read simulation data, Cython~\cite{cython_2011} to
accelerate the nanomotor stochastic simulations, gfortran~\cite{gfortran-4.9.2-doc} to build
the RMPCDMD code.

\bibliographystyle{unsrtnat}
\bibliography{/home/pierre/code/bibfile/pdebuyl}

\end{document}